\documentclass[pra]{revtex4-1}
\usepackage{graphicx}
\begin{document}
\title{Cascade Failure in a Phase Model of Power Grids}
\author{Hidetsugu Sakaguchi and Tatsuma Matsuo}
\affiliation{Department of Applied Science for Electronics and Materials,
Interdisciplinary Graduate School of Engineering Sciences, Kyushu
University, Kasuga, Fukuoka 816-8580, Japan}
\begin{abstract}
We propose a phase model to study cascade failure in power grids composed of generators and loads.  If the power demand is below a critical value, the model system of power grids maintains the standard frequency by feedback control. On the other hand, if the power demand exceeds the critical value, an electric failure occurs via step out (loss of synchronization) or voltage collapse. The two failures are incorporated as two removal rules of generator nodes and load nodes. We perform direct numerical simulation of the phase model on a scale-free network and compare the results with a mean-field approximation.
\end{abstract}
\maketitle
\maketitle
\section{Introduction}

The dynamics in complex network has been intensively studied, since the proposal of the small world network and the scale-free network.\cite{rf:1,rf:2} In the scale-free network, the degree distribution obeys a power law. 
Barab\'asi and Albert showed that a power grid in western United States was a typical example of scale-free networks.\cite{rf:2} However, there is a controversy on the degree distribution of power grids. Some authors reported  power-law distributions, whereas others reported  exponential distributions.\cite{rf:3,rf:4}  The distribution might be different for different countries and regions.  In some regions, a two-dimensional planar network might be a good approximation.  Furthermore, there is a hierarchical structure from the load dispatching center to power stations via many  transformer substations in electric power systems.   

A serious issue in power grids is a large-scale failure of power supply, called a blackout.\cite{rf:5} A large-scale blackout or a cascade failure in a power grid network can have disastrous consequences in modern society. From a viewpoint of statistical physics, several simple models have been proposed to study the cascade failure.\cite{rf:6,rf:7}  An abstract network is assumed for a power system in these models. The nodes in the network correspond to generators or consumers and the links correspond to transmission lines. Some removal rules of links in the network were assumed to simulate an elemental failure process.  A cascade failure could be reproduced by numerical simulation of some simple models, and analytical theories were proposed.

However, in actual alternating current (AC) systems, not only the network structure but also the electric potentials or phase variables are important to quantitatively describe a large-scale failure of power supply. For example, there is a phenomenon called voltage collapse, which occurs in power grids. In a voltage collapse, the electric potential at a consumer (a load) cannot be maintained at a stable stationary value and power supply to the load becomes impossible when the power demand is beyond a threshold, which can lead to a large-scale blackout.

Taking the importance of phase variables into consideration, several authors proposed coupled oscillator models similar to the Kuramoto model as a model of power grids, and synchronization transition was studied in the models.\cite{rf:8,rf:9,rf:10}  In synchronous generators, the generator turbine is driven by mechanical power, and the rotational motion is directly transformed into the sinusoidal oscillation of an alternating current.  The phase variable of the alternating current for the $i$th generator is denoted as $\theta_i$. The power produced at a generator is dissipated, accumulated, and transmitted along the electric transmission line. The output power is expressed using a sinusoidal function of the phase difference between the generators in the AC system.  A simple energy balance equation leads to a set of phase equations:
\begin{equation}
\frac{d^2\phi_i}{dt^2}=-D\frac{d\phi_i}{dt}+W_i-F\sum_j\sin(\phi_i-\phi_j),\end{equation}
where $\phi_i=\theta_i-\omega_0t$, $\omega_0$ is the standard frequency, $D$ is the damping constant, $W_i$ is related to the input power supplied to the $i$th element, and $F$ is the coupling constant.  The phase variable $\phi_i$ denotes the difference between $\theta_i$ and the phase advance $\omega_0t$ by the standard frequency.
The authors performed some numerical simulations and obtained some theoretical results.

When the damping effect is sufficiently large, the Kuramoto model is recovered as 
\begin{equation}
\frac{d\phi_i}{dt}=\omega_i-K\sum_j\sin(\phi_i-\phi_j).
\end{equation}
Phase transitions in globally coupled systems were studied in ref.~11 and ref.~12.
Recently, synchronization in scale-free networks has been studied by several authors~\cite{rf:13,rf:14,rf:15}, and a model with two subpopulations corresponding to generators and loads was also studied in ref.~16.
In these models, the mutual synchronization is a main issue of research. As a result of the mutual synchronization, a macroscopic oscillation with a certain average frequency appears.  The average frequency of the macroscopic oscillation is determined internally by self-organization in coupled oscillators, such as eqs.~(1) and (2).

However, it is important to maintain the standard frequency $\omega_0$ at a constant value, such as 50 Hz or 60 Hz, in electric power systems. If the power demand exceeds a critical value, the standard frequency cannot be maintained, and the frequency begins to depart from the standard value. This is called loss of synchronization or step out. A step out seriously affects power grids.
In this paper, we propose a new phase model of power grids, composed of sources (generators) and consumers (loads), that can simulate the occurrence of the step out phenomenon. We numerically study cascade failure in an artificially constructed networks: a square lattice and a scale-free network.  The numerical results are compared with a mean-field approximation.

\section{Phase Model of Power Grids}
Details on electric power systems, including actual generators, can be found
in textbooks on electrical engineering, e.g., ref.~17.  In this paper, we study a very simplified model that includes generators and loads, although actual power systems are more complicated. For generators, we extend the phase model described by eq.~(1).
The equation of motion for  $\phi_i$  is written as
\begin{equation}
\frac{d^2\phi_i}{dt^2}=-D\frac{d\phi_i}{dt}+W_i-P_{gi},
\end{equation}
where $D$ is the damping constant, $W_i$ is related to the input power, and $P_{gi}$ denotes the effective output power. 
Note that $d\phi_i/dt=d\theta_i/dt-\omega_0$ denotes the deviation of the frequency of the $i$th generator from the standard frequency $\omega_0$.
The output power $P_{gi}$ is expressed as
\begin{equation}
P_{gi}=\sum_j\{E_iE_jY_{ij}\sin(\phi_i-\phi_j-\alpha_{ij})+E_i^2\sin\alpha_{ij}\},
\end{equation}
where $E_i$ is the electric potential at the $i$th node, $Y_{ij}=1/\sqrt{R_{ij}^2+\omega_0^2L_{ij}^2}$, and $\alpha_{ij}={\rm tan}^{-1}\{R_{ij}/(\omega_0 L_{ij})\}$. Here $R_{ij}$ denotes the resistance between the $i$th and $j$th nodes, and $L_{ij}$ denotes the impedance between these two nodes. If $R_{ij}=0$, then $\alpha_{ij}=0$, and the sine of the phase difference $\phi_i-\phi_j$ multiplied by $E_iE_j$ contributes to the effective power. The summation in eq.~(4) is taken for all $j$ nodes, including the generators and loads linked to the $i$th generator.  In this paper, for simplicity, the output voltage $E_i$ for generators is assumed to take a constant value $E_{Gi}$.

In contrast to the models studied in ref.~8 and ref.~9, the proposed model incorporates an additional equation to keep the effective frequency $d\phi/dt$ at zero.
We use a simple feedback control system expressed as
\begin{equation}
\frac{dW_i}{dt}=-\gamma \frac{d\phi}{dt}.
\end{equation}
This feedback control works as follows. If $d\phi/dt$ is positive or the effective frequency is too fast, the input power is decreased. Conversely, if $d\phi/dt$ is negative or the effective frequency is too slow, the input power is increased. If the feedback control works well, the effective frequency $d\phi_i/dt$  is maintained at zero. Under this condition, the entire power system, which includes the generators and loads, is synchronized at the standard frequency $\omega_0$, such as 50 Hz or 60 Hz. 
In actual power generators, the frequency is controlled by a local feedback using governors and a global regulation by the control center. Equation (5) corresponds to the local feedback by governors. Equation (5) implies that $W_i(t)+\gamma\phi_i(t)=$const. If the constant is set to zero, then $W_i=-\gamma\phi_i$. Then, eq.~(3) takes a form similar to that of the equation of motion of a damping harmonic oscillator with an output power term.
\begin{equation}
\frac{d^2\phi_i}{dt^2}=-D\frac{d\phi_i}{dt}-\gamma\phi_i-P_{gi}.
\end{equation}
This is our model equation for generators. The form of the equation is similar to that of eq.~(1); however, the dynamical behavior is rather different from that in the case of eq.~(1).

If the power demand expressed by $P_{gi}$ is too large, the input power $W_i$ cannot respond to the excessive demand. If the generators become overloaded, they might break down.  To avoid a breakdown, the generators might be stopped. In such an event, the overall synchronization would be lost; that is, a loss of synchronization or step out would occur, which might lead to a cascade failure.
 To express such a situation, we assume a simple rule that $E_i$ is set to zero if $W_i$ is beyond a critical value $W_{ci}$. If $E_i(t)$ is set to zero, the function of the generator is lost. It is equivalent to effectively removing the generator node $i$ from the power network system.

For loads, on the other hand, we use different energy balance equations. That is, we assume that definite quantities of effective power $P_{ei}$ and reactive power $Q_{ri}$ are consumed at each load $i$. The complex power $P_{ei}+iQ_{ri}$ is expressed as $E_{i}e^{i\phi_i}I_i^{*}$, where $I_i$ is the current supplied to the electric devices at the $i$th node, and $^*$ implies the complex conjugate.  The current $I_i$ is expressed as
\[I_i=\sum_{j}\frac{E_je^{i\phi_j}-E_ie^{i\phi_i}}{R_{ij}+i\omega_0 L_{ij}}=\sum_j-iY_{ij}e^{i\alpha_{ij}}\left (E_je^{i\phi_j}-E_ie^{i\phi_i}\right )\]
using the difference between the electric potentials and the admittance. Here, $1/(R_{ij}+i\omega L_{ij})$ is denoted as $-iY_{ij}e^{i\alpha_{ij}}$.   The powers $P_{ei}$ and $Q_{ri}$ are therefore expressed as
\begin{eqnarray}
P_{ei}&=&\sum_jY_{ij}\{E_iE_j\sin(\phi_j-\phi_i+\alpha_{ij})-E_i^2\sin\alpha_{ij}\},\nonumber\\
Q_{ri}&=&\sum_jY_{ij}\{E_iE_j\cos(\phi_j-\phi_i+\alpha_{ij})-E_i^2\cos\alpha_{ij}\}.
\end{eqnarray}
 The summation is taken for all sites $j$, including the generators and loads linked to the $i$th node.

Equations (7) are coupled algebraic equations. The phase $\phi_i$ and electric potential $E_i$ are determined for each $i$ by solving eq.~(7) for specified values of $P_{ei}$ and $Q_{ri}$.
We have numerically solved the coupled equations (7) with an iterative method.
Generally, if $P_{ei}$ and $Q_{ri}$ are sufficiently small, both stable and unstable stationary solutions exist for $E_i$. However, if the effective power $P_{ei}$ is gradually increased as a control parameter, the magnitude of a stable solution $E_i$ decreases and that of an unstable one increases.  At a critical value of $P_{ei}$, a transition occurs such that the stable and unstable solutions merge and disappear. There are no stationary solutions  beyond the critical value.  This phenomenon is called voltage collapse. If a voltage collapse occurs, the voltage $E_i$ decreases rapidly, which can lead to an electric failure of the power supply.
 A voltage collapse occurs more easily if the reactive power $Q_{ri}$ is large.  If a voltage collapse occurs, there is no stationary solution in the coupled equations (7). In our simple model, we have assumed another rule that $E_i$ is set to zero when $E_i(t)$ decreases to zero.  This operation is equivalent to removing the load node $i$ from the network of power grids, because $E_i$ appears as a coefficient for the interaction terms in eq.~(7).

Thus, we have proposed a new simple phase model of power grids with two removal rules. The two removal rules simulate the situations of step out and the voltage collapse. Our model is similar to the models in ref.~6 and ref.~7 in that some removal rules are assumed. At the same time, our phase model has a form similar to that of eq.~(1) studied in ref.~8 and ref.~9, although the dynamical behavior is rather different because feedback control is very important in our model. We perform numerical simulations of coupled equations (6) and (7) on a scale-free network for the sake of simplicity, although real power-grids are not regular square lattice and a scale-free network. Our model systems can be easily applied to other types of networks. Actually, we have performed numerical simulation on a square lattice and found cascade failure. 
 The scale-free network is constructed using the preferential attachment method of Barab\'asi and Albert.  New nodes are randomly assigned as being generators with probability $p$ or loads with probability $1-p$. 

\section{Numerical Simulations on  a Scale-Free Network}
In this paper, we show some numerical results for a uniform system where
the parameters do not depend on $i$, i.e., $P_{ei}=P_e,Q_{ri}=Q_r,E_{Gi}=E_G,W_{ci}=W_c, Y_{ij}=Y$, and $\alpha_{ij}=\alpha$. The total number of nodes is denoted by $N$. Some parameters
 in this paper are fixed, namely, $D=1,\gamma=1, Q_r=0.001,Y=1$, and $E_G=1$, whereas $N,W_c, P_e,p$ and $\alpha$ are changed as control parameters.

Figure 1 shows a numerical result of a scale-free network of $N=100$ at $p=0.15,P_e=0.2, W_c=1.5$, and $\alpha=0$. The initial conditions are $\phi_i=0,d\phi_i/dt=0,E_i=1$, and $W_i=0$. There are 15 generators and 85 loads in this network. Figure 1(a) shows a network at $t=2$, where no failure occurs. The rhombi denote generator nodes and the plus signs denote load nodes. Figure 1(b) shows a network at $t=10$, which is a stationary state. The input power $W_i$ of one generator $i=6$ of degree 14 goes beyond the critical value $W_c$, and a loss of synchronization occurs; hence, the voltage $E_i$ is set to zero.
The generator node $i=6$ is one of the hubs with large degree. The power demand to the other generators increases by the effective removal of this generator.
As a result, a loss of synchronization occurs successively at two other generator nodes, namely, $i=1$ of degree 24 and $i=9$ of degree 13.  Meanwhile, as a result of the shortage of the power supply, a large-scale voltage collapse occurs for the load nodes. When the voltage collapse occurs, the load nodes are effectively removed. Then, the successive loss of synchronization stops, because power demand to the other generators decreases.  Finally, 12 generators and 5 loads survive in this numerical simulation, as shown in Fig.~1(b). As the numerical simulation suggests, the loss of synchronization tends to occur like a
cascade failure; however, the voltage collapse tends to suppress the cascade failure because the power demand decreases when the load nodes are removed.
\begin{figure}[t]
\begin{center}
\includegraphics[height=4.cm]{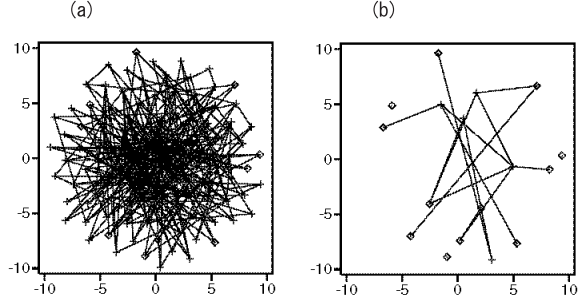}
\end{center}
\caption{(a) Scale-free network at $t=2$. Generators are denoted by rhombi and loads are denoted by $+$. (b) Network at $t=10$. The generator nodes that exhibit step out and the load nodes that exhibit voltage collapse are removed; moreover, the links leaving the removed nodes are also removed.
}
\label{f1}
\end{figure}

\section{Mean-Field Approximation}
Our system can be simplified with a mean-field approximation. The mean-field approximation is considered to be good if the degree of each node is sufficiently large.   A similar mean-field analysis was performed for coupled phase oscillators in a scale-free network.\cite{rf:14,rf:15}  Note that
$D=\gamma=1$ is assumed in the following equations. The phases $\phi_i$'s of generators and loads at nodes of degree $l$ are assumed to take the same value and are respectively denoted as $\phi_{Gl}$ and $\phi_{Ll}$. Similarly, the voltages $E_i$'s at generator nodes and load nodes of degree $l$ are assumed to take the same value and are respectively denoted as $E_{Gl}$ and $E_{Ll}$.
The voltage $E_{Gl}$ takes a constant value $E_G$ when the generator nodes do not exhibit step out, and it becomes zero when the generator nodes exhibit step out. On the other hand, $E_{Ll}$'s are determined by coupled equations (10) and (11) and become zero if the loads exhibit voltage collapse.  The probability distribution of degree $l$ is denoted as $P(l)$. The distribution $P(l)$ obeys a power law in a scale-free network, but our theory can be applied to any degree distribution $P(l)$. We have used the distribution $P(l)$ of the randomly constructed network used in our numerical simulation.
$|Y|$ is denoted as $Y$. We introduce two kinds of order parameters, $\sigma_G$ and $\sigma_L$, for generators and loads, respectively.
\begin{eqnarray}
\sigma_G\exp(i\phi_G)&=&\frac{\sum P(l)lE_{Gl}e^{i\phi_{Gl}}}{\sum P(l)l},\nonumber\\
\sigma_L\exp(i\phi_L)&=&\frac{\sum P(l)lE_{Ll}e^{i\phi_{Ll}}}{\sum P(l)l}.
\end{eqnarray}
They are average values of $E_{Gl}e^{i\phi_{Gl}}$ and $E_{Ll}e^{i\phi_{Ll}}$ weighted with degree $l$.
By using these order parameters, eq.~(6) is approximated as
\begin{eqnarray}
\frac{d^2\phi_{Gl}}{dt^2}&=&-\phi_{Gl}-\frac{d\phi_{Gl}}{dt}+plYE_{Gl}\{\sigma_G\sin(\phi_G-\phi_{Gl}+\alpha)-E_{Gl}\sin\alpha\}\nonumber\\
& &+(1-p)lYE_{Gl}\{\sigma_L\sin(\phi_L-\phi_{Gl}+\alpha)-E_{Gl}\sin\alpha\}.
\end{eqnarray}
On the other hand, eq.~(7) is approximated as
\begin{eqnarray}
P_e&=&plYE_{Ll}\{\sigma_G\sin(\phi_G-\phi_{Ll}+\alpha)-E_{Ll}\sin\alpha\}\nonumber\\
& &+(1-p)lYE_{Ll}\{\sigma_L\sin(\phi_L-\phi_{Ll}+\alpha)-E_{Ll}\sin\alpha\},\\
Q_r&=&plYE_{Ll}\{\sigma_G\cos(\phi_G-\phi_{Ll}+\alpha)-E_{Ll}\cos\alpha\}\nonumber\\
& &+(1-p)lYE_{Ll}\{\sigma_L\cos(\phi_L-\phi_{Ll}+\alpha)-E_{Ll}\cos\alpha\}.\nonumber\\
\end{eqnarray}
As stated already, the additional rule for step out is that $E_{Gl}$ is set to zero when $W_l=-\phi_{Gl}$ exceeds the critical value $W_c$. The other additional rule for voltage collapse is that $E_{Ll}$ is set to zero when there are no stationary solutions to eqs.~(10) and (11).
We have performed numerical simulation of coupled equations (8), (9), (10), and (11) to find a solution in the mean-field theory.

\begin{figure}[t]
\begin{center}
\includegraphics[height=4.cm]{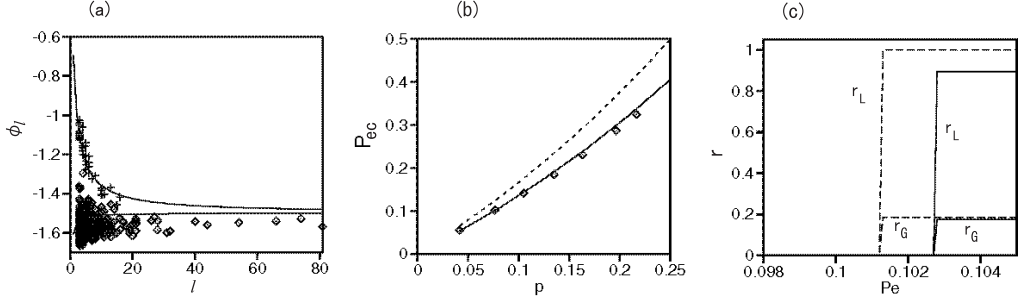}
\end{center}
\caption{(a) Relation between the phase $\phi_i$ and the degree $l$ in the scale-free network. The $+$ signs and rhombi denote numerical results for generators and loads. Solid curves denote numerical results for the mean-field equation. The upper one corresponds to generators and the lower one corresponds to loads. (b) Critical values $P_{ec}$ as functions of the number ratio $p$ of generators. The rhombi denote the results of the direct numerical simulation. The solid curve is the result obtained using the mean-field equation, and the dotted curve denotes eq.~(17). (c) Number ratios $r_G$ and $r_L$ of removed nodes as functions of $P_e$ for $p=0.077$.}
\label{f3}
\end{figure}
Figure 2(a) compares the result for a scale-free network of $N=600$ at $\alpha=0,P_e=0.1$, and $p=46/600=0.077$ obtained by direct numerical simulation with that obtained by mean-field approximation. The entire system is synchronized, and voltage collapse does not occur at these parameter values. The horizontal axis denotes degree $l$, and the vertical axis denotes $\phi_i$ at the stationary state. The phases $\phi_i$'s for generators are denoted by $+$, and the $\phi$'s for loads are denoted by rhombi. The two solid curves represent $\phi_{Gl}$ and $\phi_{Ll}$ obtained by numerical simulation of the mean-field equation.  Rather good agreement is observed. As $P_e$ is increased, loss of synchronization occurs first, when $W_i=-\phi_i$ for a generator with a large degree $l$ exceeds the critical value $W_c=1.5$.  The critical values $P_{ec}$ of $P_{e}$ for step out have been calculated as functions of $p$ by direct numerical simulations of $N=600$ and $\alpha=0$.  The results are shown with rhombi in Fig.~2(b). The numerical result obtained using the mean-field equation is shown by a solid curve. Fairly good agreement is observed.
It is natural that the critical value of the effective power increases with the number of the generators.

The mean-field equation is still rather complicated. If we further assume that $\sigma_G\sim 1,\sigma_L\sim 1,\alpha=0,E_{Gl}=1,E_{Ll}\sim 1,Y=1$, and $|\phi_{Gl}-\phi_{Ll}|<<1$ (which is approximately the case for small $p$), the critical value of step out can be approximately evaluated.
The stationary solution of eq.~(9) satisfies
\begin{equation}
-\phi_{Gl}+pl(\phi_G-\phi_{Gl})+(1-p)l(\phi_L-\phi_{Gl})=0.
\end{equation}
That is,
\begin{equation}
\phi_{Gl}=\frac{pl\phi_G+(1-p)l\phi_L}{1+l}\sim\{p\phi_G+(1-p)\phi_L\}(1-1/l).
\end{equation}
Similarly, eq.~(10) yields
\begin{equation}
\phi_{Ll}=\frac{pl\phi_G+(1-p)l\phi_L-P_e}{l}.
\end{equation}
Substitution of eqs.~(13) and (14) into eq.~(8) yields
\begin{eqnarray}
\exp\{i(1-p)(\phi_G-\phi_L)\}&=&\frac{\sum lP(l)e^{-i\{p\phi_G+(1-p)\phi_L\}/l}}{\sum lP(l)},\nonumber\\
\exp\{ip(\phi_L-\phi_G)\}&=&\frac{\sum lP(l)e^{-iP_e/l}}{\sum lP(l)}.
\end{eqnarray}
By using the approximation $\exp(ix)\sim 1+ix$, the following approximate relations are obtained:
\begin{eqnarray}
\phi_G-\phi_L&=&P_e/(p\langle l\rangle), \nonumber\\
p\phi_G+(1-p)\phi_L&=&-(1-p)P_e/p,\nonumber\\
\phi_{Gl}&=&-(1-p)P_e/p(1-1/l),\nonumber\\
\phi_{Ll}&=&-(1-p)P_e/p-P_e/l,
\end{eqnarray}
where $\langle l\rangle$ is the average value of degree and is given by $\langle l\rangle=\int P(l)ldl$. As is seen from eq.~(16),
$\phi_{Gl}$ decreases as $1/l$ and $\phi_{Ll}$ increases as $-1/l$ and $\phi_{Gl}$ and $\phi_{Ll}$ approaches for $l\rightarrow \infty$. 
The numerical results for the mean-field equation shown by the solid curves in Fig.~2(a) exhibit similar behavior that $\phi_{Gl}$ decreases as $1/l$ and $\phi_{Ll}$ increases as $-1/l$ for large $l$.  
The maximum value of $W_l$ is evaluated as $-\phi_{Gl}=(1-p)P_e/p(1-1/l)\sim (1-p)P_e/p$. Therefore, the critical value of $P_e$ is evaluated as
\begin{equation}
P_{ec}=\frac{pW_c}{1-p}.
\end{equation}
The dotted curve in Fig.~3(b) denotes eq.~(17).
The rough approximation seems to be sufficiently accurate for small $p$.

 We have numerically calculated $r_G$ and $r_L$ also for the scale-free network of $N=600$ . Figure 2(c) shows $r_G$ and $r_L$ as functions of $P_e$ for $p=0.077$. The solid curves denote results of the direct numerical simulation and the dashed curves denote results of the mean-field model. It is seen that the transitions are discontinuous, and the mean-field approximation is fairly good. 

\section{Summary}
We have proposed a phase model of power grids.
A simple feedback control was used to maintain the standard frequency in the model. Electric failure occurred via voltage collapse and step out (loss of synchronization).  The two electric failures were incorporated as two removal rules for nodes of generators and loads. We performed direct numerical simulation and studied the mean-field model on a square lattice and in a scale-free network. 
We have found two kinds of transitions in a large network including many generators and loads.  The step-out dominant failure expands as a cascade failure.  
The mean-field approximation was fairly suitable for understanding the complicated behavior of this system.  In this paper, we used a regular square lattice and the scale-free network of Barab\'asi and Albert, and we assumed a uniform system for the sake of simplicity.  Our model can be applied to general networks, and the mean-field theory might be applied to dense networks with large degrees. We have assumed that the critical value of power supply $W_{ci}$ and the power demand $P_{ei}$ are the same for all $i$'s; however, in practice, these values are generally distributed widely. It is left to future research to study more realistic power grids with heterogeneous power supply and demand.

\end{document}